\documentclass[5p,twocolumn,number,times]{elsarticle}

\usepackage{graphics}
\usepackage{amssymb}
\journal{Nuclear Instruments and Methods in Physics Research, Section A}

\begin{document}

\begin{frontmatter}



\title{Radio detection of cosmic rays in the Pierre Auger Observatory}


\author[FZK]{T. Huege}
\author[Coll]{for the Pierre Auger Collaboration}

\address[FZK]{Institut f\"ur Kernphysik, Forschungszentrum Karlsruhe, Postfach 3640, 76021 Karlsruhe, Germany}
\address[Coll]{Av. San Martin Norte 304, (5613) Malarg\"ue, Prov. de Mendoza, Argentina}

\begin{abstract}
In small-scale experiments such as CODALEMA and LOPES, radio detection of cosmic rays has demonstrated its potential as a technique for cosmic ray measurements up to the highest energies. Radio detection promises measurements with high duty-cycle, allows a direction reconstruction with very good angular resolution, and provides complementary information on energy and nature of the cosmic ray primaries with respect to particle detectors at ground and fluorescence telescopes. Within the Pierre Auger Observatory, we tackle the technological and scientific challenges for an application of the radio detection technique on large scales. Here, we report on the results obtained so far using the Southern Auger site and the plans for an engineering array of radio detectors covering an area of $\sim$20 km$^{2}$.
\end{abstract}

\begin{keyword}
cosmic rays \sep extensive air showers \sep electromagnetic radiation from moving charges

\PACS 96.50.S- \sep 96.50.sd \sep 41.60.-m


\end{keyword}

\end{frontmatter}


\section{Introduction}

Our understanding of the highest-energy cosmic rays, studied in detail with the Pierre Auger Observatory \citep{AugerNIM2004}, is still rather limited. One of the reasons is that their event rates are extremely low --- less than one particle per km$^{2}$ and century at the highest energies. The Pierre Auger Observatory, currently comprising the Southern site in Argentina, observes these highest energy cosmic rays with a combination of surface particle detectors (SD) and fluorescence telescopes (FD). This ``hybrid'' mode of observation provides a wealth of information on each single measured cosmic ray event and is the key for the successful interpretation of the measured cosmic ray data. It is, however, limited to a duty cycle of $\sim 13$\% by the fluorescence technique, which can only be applied during clear, moonless nights.

A long-forgotten detection technique based on the measurement of pulsed radio emission emanating from cosmic ray air showers has great potential to complement the existing techniques of particle and air fluorescence detection. Radio detection of cosmic ray air showers was originally discovered in the 1960s \citep{JelleyFruinPorter1965}, but was nearly forgotten after the 1970s and only revived with great success in the last few years. Radio pulses with a length of a few tens of nanoseconds are emitted by relativistic secondary electrons and positrons in the air shower cascade which ensues the interaction of high-energy cosmic rays with atmospheric nuclei. The acceleration of these electrons and positrons in the earth's magnetic field is by now known to contribute the dominant fraction of the radio emission. This geomagnetic effect has been modeled with a number of approaches (for a review see \citep{HuegeArena2008}), most prominently the geosynchrotron model \citep{HuegeFalcke2003a} and the macroscopic transverse current model \citep{ScholtenWernerRusydi2008}. One of the main advantages of the technique is that radio detection of cosmic ray air showers can be carried out with nearly 100\% duty cycle --- allowing hybrid observations of radio and surface particle detectors with statistics larger by nearly a factor of ten as compared to hybrid measurements of particle and air fluorescence detectors.

In recent years, radio detection has matured and shown promising results achieved by the CODALEMA \citep{ArdouinBelletoileCharrier2005} and LOPES \citep{FalckeNature2005} experiments. We now know (see \citep{LautridouArena2008,HaungsArena2008} and references therein) that the radio signal field strength increases linearly with the energy of the primary particle in the frequency regime of coherent emission (below $\sim$100~MHz), falls off approximately exponentially with lateral distance from the shower axis and correlates strongly with the angle between shower axis and geomagnetic field. Due to the relatively small area of these first-generation experiments, covering less than 0.5~km$^{2}$ each, radio emission of cosmic rays has only been studied up to energies of below 10$^{18}$~eV. The true potential of the radio technique, however, lies in its large scale application for the measurement of ultra-high energy cosmic rays. The combination of almost 100$\%$ duty cycle, calorimetric measurement of the air shower energy \citep{HuegeUlrichEngel2008} and sensitivity on the depth of the air shower maximum (and thus mass of the primary cosmic ray) \citep{HuegeUlrichEngel2008} makes it an ideal complement for the Auger surface particle detectors. In addition, radio detection can provide a very high angular resolution, and is believed to be cost-effective for instrumentation of large areas.

\section{Towards large scale application}

The potential of the radio detection technique for application on large scales is being evaluated in the framework of the Pierre Auger Observatory. This allows direct comparisons of radio measurements with high-quality measurements of the SD and FD detectors, which are extremely helpful in the research and development phase. In addition, the Argentinian pampa provides a very good environment for radio detection, as the radio-noise levels are very low.

A number of technological challenges have to be solved for the application of radio detection on large scales. A large array, in contrast to the smaller first-generation experiments, needs in particular:

\begin{itemize}
\item{autonomous, self-powered detector stations}
\item{communication between stations through wireless data links}
\item{ability to self-trigger on the radio signal}
\item{stations withstanding roaming animals and strong winds}
\item{long-term operational stability with minimal maintenance}
\end{itemize}

\begin{figure}[!ht]
   \centering
   \includegraphics[width=\columnwidth]{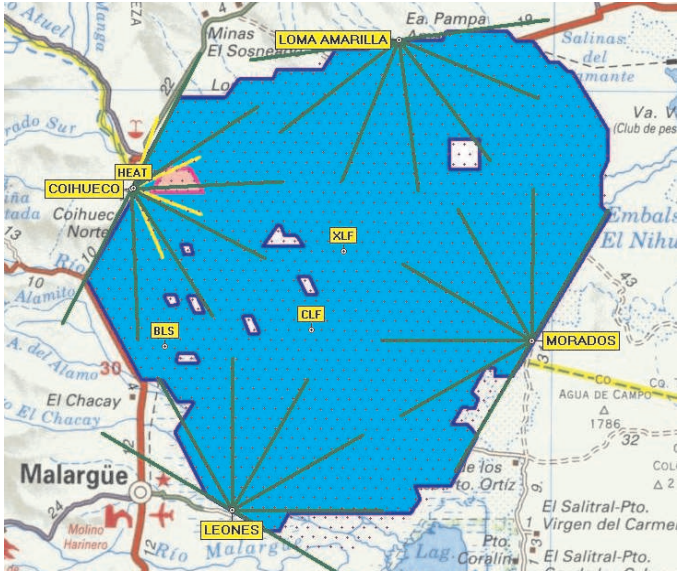}
   \caption{
   \label{augerlayout}
   Layout of the Pierre Auger Observatory. Radio prototype experiments are located near the BLS and CLF. AERA will be set up in the north-west area near Coihueco in the region marked in red.
   }
   \end{figure}

In a first phase of research and development, a number of concepts for detector stations have been studied with small prototype installations of a handful of antennas each. With installations near the Auger Balloon Launching Station (BLS, see Fig.\ \ref{augerlayout}) and Central Laser Facility (CLF, see Fig.\ \ref{augerlayout}), various aspects of the detector design have been evaluated.

%______________________________________________________________
%______________________________________________________________

\begin{figure}[!ht]
   \centering
   \includegraphics[width=\columnwidth]{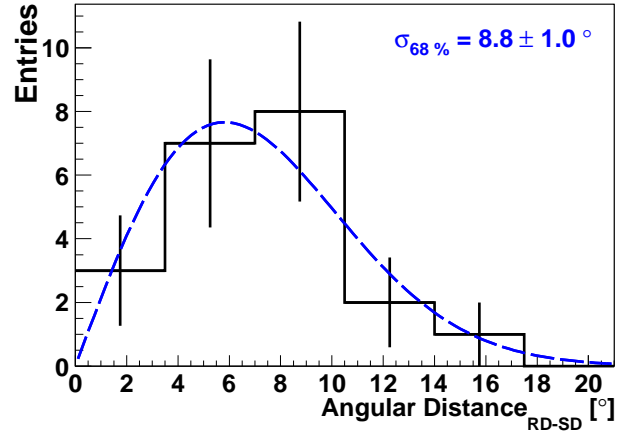}
   \caption{
   \label{angulardistance}
   Angular difference between the reconstructed arrival directions of 27 air showers observed with both the Auger surface detectors and the radio detector prototype stations near the BLS. The dashed line is a fit through the data using a Rayleigh function.
   }
   \end{figure}

\begin{figure}[!ht]
   \centering
   \includegraphics[width=0.85\columnwidth]{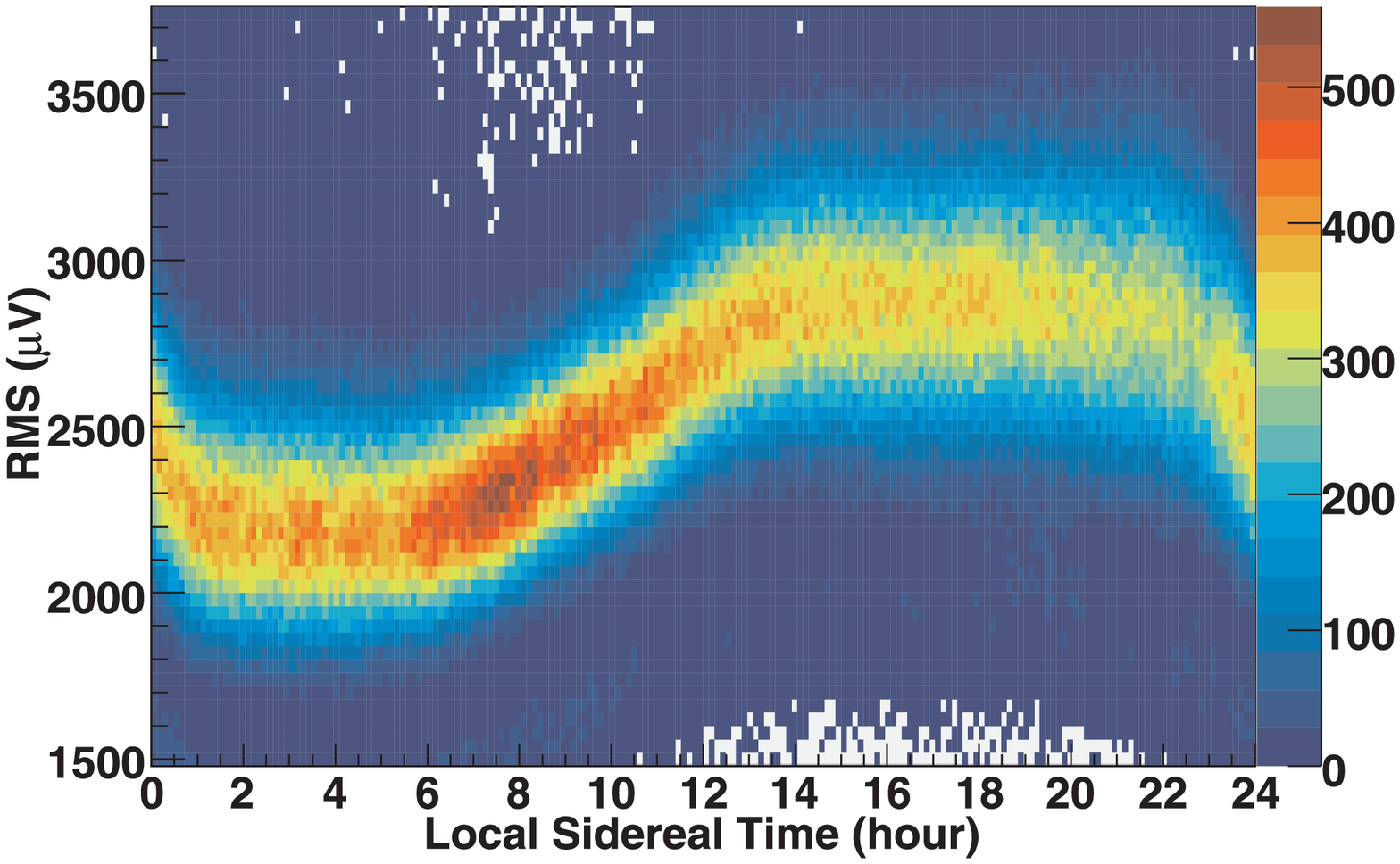}
   \includegraphics[width=0.85\columnwidth]{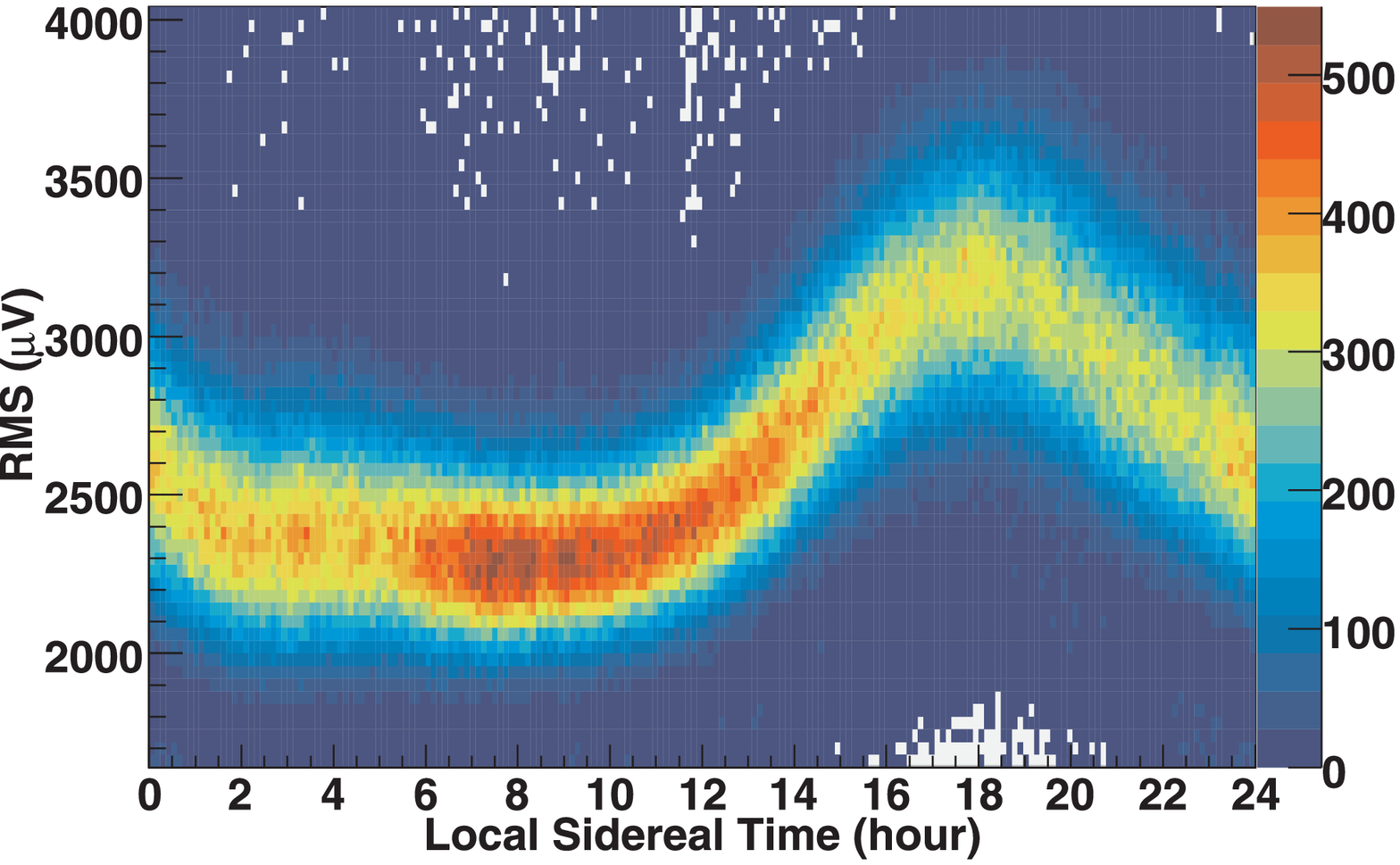}
   \caption{
   \label{noise}
   Noise level (arbitrary units) in the frequency window from 50 to 55 MHz measured by a north-south-polarised (top) and an east-west-polarised (bottom) radio antenna near the BLS as a function of local siderial time. The data cover approximately one year of measurements. The passage of the Galactic centre can be clearly identified.
   }
   \end{figure}

A cabled installation near the BLS has taken radio data trigger externally by a pair of scintillators \citep{CoppensArena2008}. Over a year of operation, $>\!25$ events have been coincidently detected in all three radio antennas, for which a direction reconstruction confirms that they correspond to events seen by the Auger surface particle detectors at the same time and from the same arrival direction (see Fig.\ \ref{angulardistance}). In addition, a full radio event and radio detector simulation has been carried out, confirming that the field strengths measured are compatible with those expected within the relatively large uncertainties \citep{FliescherArena2008}. It has also been demonstrated that the radio detector system is sensitive to galactic noise and in fact sees the passage of the Galactic centre (cf. \citep{CoppensArena2008} and see Fig. \ref{noise}), illustrating that sky noise can be used for calibration purposes.

%______________________________________________________________
%______________________________________________________________

%______________________________________________________________
%______________________________________________________________

\begin{figure}[!ht]
   \centering
   \includegraphics[width=\columnwidth]{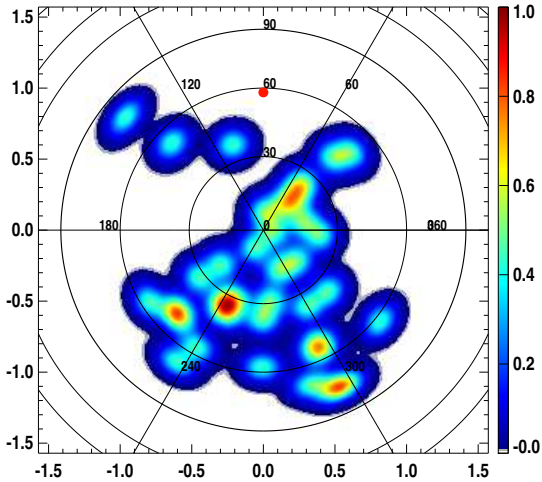}
   \caption{
   \label{skymap}
   Sky map of the 36 radio events registered by the self-triggering radio detector prototype near the CLF in coincidence with Auger surface detector events. The plot is made in local spherical coordinates with the zenith at the centre, east at 0$^{\circ}$ and north at 90$^{\circ}$ azimuth. The geomagnetic field direction in Malarg\"ue is indicated by the red point.
   }
   \end{figure}

Another installation set up near the CLF has successfully tested autonomous stations working in a radio-self-triggered mode \citep{RevenuArena2008}. Over a year of operation, 36 events have been triggered by the radio detector itself, which could afterwards be associated to events also seen by the Pierre Auger surface particle detectors. The angular distribution of these radio-triggered events shows the expected north-south asymmetry caused by the dependence of the radio field strength on the angle between air shower axis and geomagnetic field (see Fig.\ \ref{skymap}).

%______________________________________________________________
%______________________________________________________________

\begin{figure}[!ht]
   \centering
   \includegraphics[width=\columnwidth]{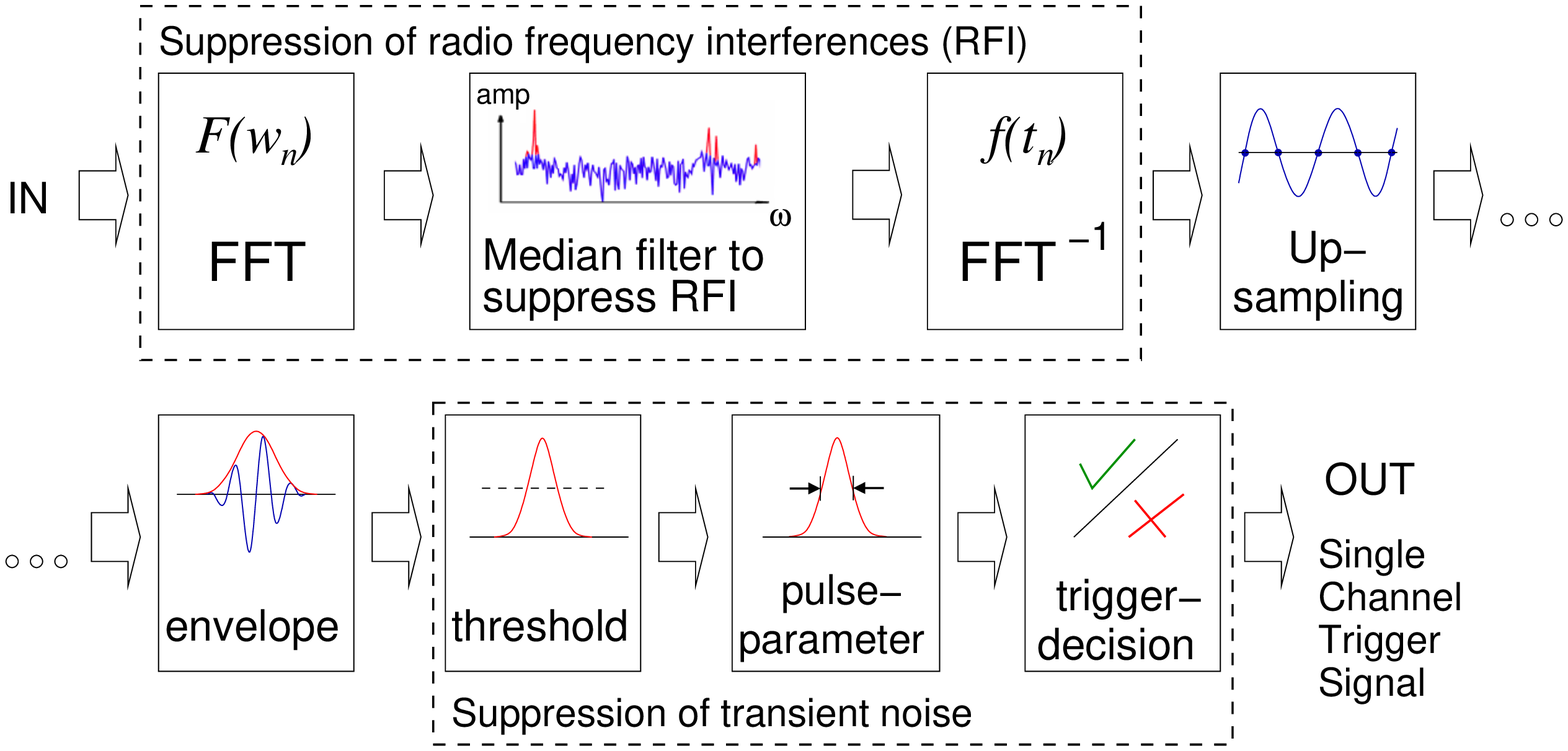}
   \caption{
   \label{trigger}
   Schematic diagram of the FPGA-based smart self-trigger algorithm being studied near the BLS (see text).
   }
   \end{figure}

In addition, further tests of fully autonomous, self-triggered stations on larger baselines are currently being carried out near the BLS \citep{CoppensArena2008}, accompanied by tests of a sophisticated FPGA-based self-trigger which --- in real-time --- performs suppression of narrow-band radio-frequency interference, upsamples and envelopes the RF-pulses, determines important pulse parameters such as signal-to-noise ratio, pulse width and pulse multiplicity, and then makes a trigger decision on these parameters to suppress impulsive noise but retain cosmic ray candidates (see Fig.\ \ref{trigger}).

%______________________________________________________________
%______________________________________________________________

\section{The Auger Engineering Radio Array}

%______________________________________________________________
   \begin{figure*}[!ht]
   \centering
   \includegraphics[width=\textwidth]{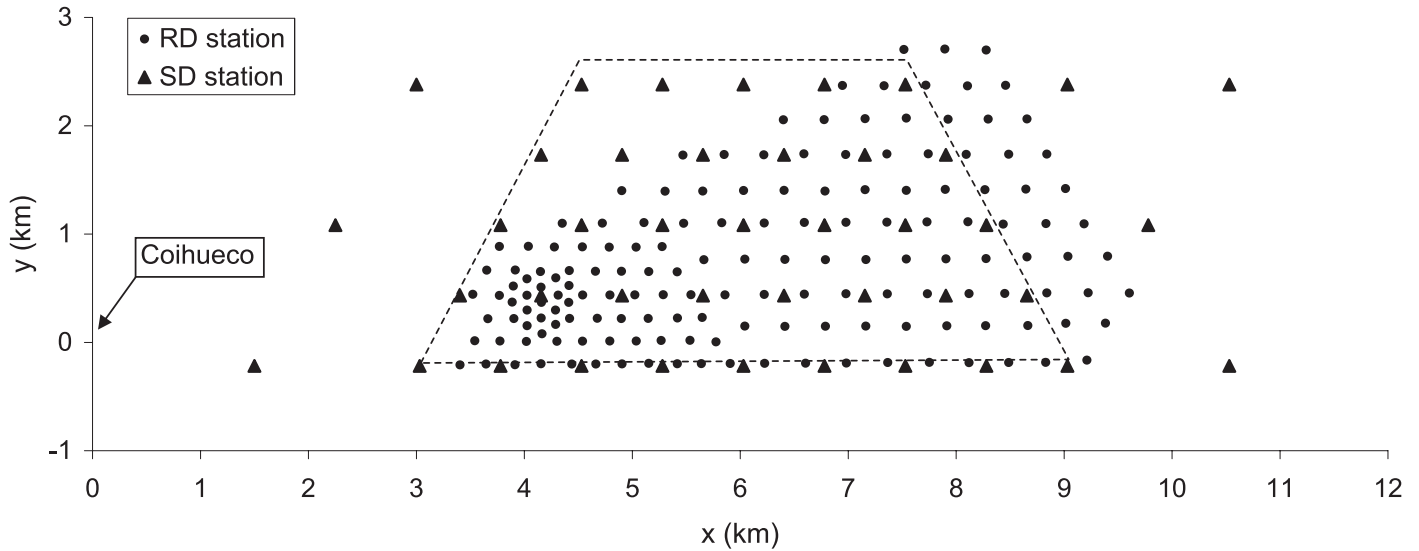}
   \caption{
   \label{aeralayout}
   Layout of the planned Auger Engineering Radio Array (AERA). Radio detector stations will be put on triangular grids with grid constants of 175~m, 250~m and 375~m. Also indicated are the upper half of the hexagonal area foreseen for infill surface detectors of AMIGA and the location of the Coihueco fluorescence detector telescopes.
   }
   \end{figure*}
%______________________________________________________________

Based on the promising results of the initial prototype installations, the Pierre Auger Collaboration has decided to set up a larger scale Auger Engineering Radio Array (AERA) \citep{vandenBergIcrc2009}. In its final form, AERA will cover an area of $\sim 20$~km$^{2}$ with approximately 150 detector stations, and will detect about 5000 cosmic ray radio events per year. It will be set up in the north-western part of the observatory (marked red in Fig.\ \ref{augerlayout}), co-located with the Auger enhancements HEAT and AMIGA. The co-location with HEAT (high elevation fluorescence telescopes, see \citep{KlagesICRC2007}) and AMIGA (an array of infill surface detectors and underground muon detectors, see \citep{EtchegoyenIcrc2007}) will provide ample data for hybrid data analysis in the energy range covered by AERA.

AERA will serve as {\em the} prototype for a large scale radio detector array, and will address, in the given order, the following science goals:

\begin{enumerate}
\item{Thorough analysis of the radio emission from cosmic ray air showers at energies above $\sim 10^{17.2}$~eV. Thereby, we will verify if the dependences of radio emission on air shower parameters at high energies are identical to those measured at lower energies, as predicted by various theoretical models \citep{HuegeArena2008}.}
\item{Evaluation of the capability of large-scale radio detection to provide observables important for cosmic ray measurements, in particular the energy of the primary particle, mass of the primary particle, and arrival direction of the primary cosmic ray.}
\item{Highly competitive measurements of the energy spectrum and cosmic ray mass composition in the region of transition from galactic to extragalactic sources (i.e., between $\sim 10^{17.4}$ and $\sim 10^{18.7}$~eV).}
\end{enumerate}

To reach these goals, AERA will comprise a total of $\sim 150$ antennas set up in a multi-scalar layout (see Fig.\ \ref{aeralayout}) with a dense core of 24 antennas spaced on a triangular grid of 150~m, encompassed by 60 antennas on a triangular grid of 250~m and an outer region of 72 antennas on a triangular grid of 375~m. Each of the autonomous detector stations will be optimized for an observing frequency window from 30 to 80~MHz, perform self-triggering on radio pulses with an FPGA-based smart trigger, and sample the data with 4 channels of 12 bit ADCs at a sampling rate of 200 MS~s$^{-1}$. To allow read-out of detector stations which have not self-triggered, a ring-buffer corresponding to approximately 3 seconds of data is foreseen.

%______________________________________________________________
   \begin{figure}[!htb]
   \centering
   \includegraphics[width=0.85\columnwidth]{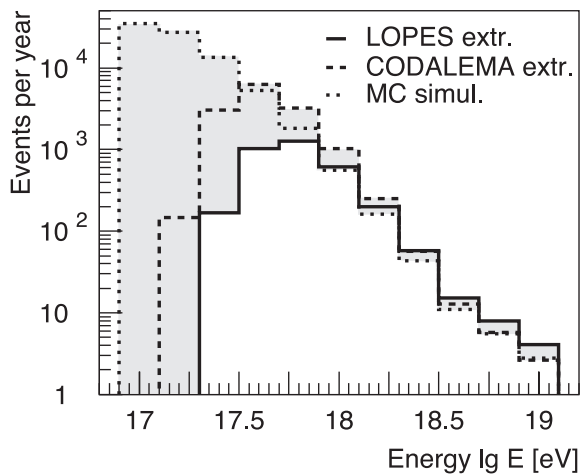}
   \caption{
   \label{rates}
   Projected event rates for AERA based on extrapolations of LOPES data, CODALEMA data and simulations with the REAS2 radio emission code.
   }
   \end{figure}
%______________________________________________________________

Event rates expected for AERA have been calculated on the basis of CODALEMA data, LOPES data and simulations using the REAS2 code \citep{HuegeUlrichEngel2007a} and are depicted in Fig.\ \ref{rates}. Once the array is complete, $\sim 5000$ events will be measured by AERA per year, $\sim 1000$ of which with energies above 10$^{18}$~eV. From conservative estimates, the lower energy threshold of AERA will be at $\sim 10^{17.2}$~eV.

\section{Summary}

Over the last few years, radio detection of cosmic rays has matured and is now at the verge of large-scale application. Initial prototype radio measurements within the Pierre Auger Observatory have been successful. As a logical next step, the 20~km$^{2}$ Auger Engineering Radio Array (AERA) will be set up. AERA will provide us with a detailed understanding of the radio emission physics at energies above $\sim 10^{17.5}$~eV, evaluate the potential of the radio detection technique for application on scales as large as the Pierre Auger Observatory, and contribute to the study of the transition of cosmic rays from galactic to extragalactic sources with highly competitive measurements.



%\bibliography{tims_references}

\begin{thebibliography}{10}

\expandafter\ifx\csname url\endcsname\relax
  \def\url#1{\texttt{#1}}\fi
\expandafter\ifx\csname urlprefix\endcsname\relax\def\urlprefix{URL }\fi
\expandafter\ifx\csname href\endcsname\relax
  \def\href#1#2{#2} \def\path#1{#1}\fi

\bibitem{AugerNIM2004}
J.~{Abraham}, M.~{Aglietta}, I.~C. {Aguirre}, {et al. (The Pierre Auger Collaboration)},
%{Properties and performance of the prototype instrument for the Pierre Auger Observatory},
  Nucl. Instrum. Meth. A523 (2004) 50--95.
  \newblock \href {http://dx.doi.org/10.1016/j.nima.2003.12.012}
  {\path{doi:10.1016/j.nima.2003.12.012}}.

\bibitem{JelleyFruinPorter1965}
J.~V. {Jelley}, J.~H. {Fruin}, N.~A. {Porter}, {et al.}, Nature 205 (1965) 327.

\bibitem{HuegeArena2008}
T.~{Huege},
%{Simulations and theory of radio emission from cosmic ray air showers},
  in: Proc. of the ARENA 2008 Conference, Rome, Italy, NIM A, 2009.
  \newblock \href {http://dx.doi.org/10.1016/j.nima.2009.03.165}
  {\path{doi:10.1016/j.nima.2009.03.165}}.

\bibitem{HuegeFalcke2003a}
T.~{Huege}, H.~{Falcke}, Astronomy \& Astrophysics 412 (2003) 19--34.
  \newblock \href {http://dx.doi.org/10.1051/0004-6361:20031422}
  {\path{doi:10.1051/0004-6361:20031422}}.

\bibitem{ScholtenWernerRusydi2008}
O.~{Scholten}, K.~{Werner}, F.~{Rusydi}, Astropart. Physics 29
  (2008) 94--103.
  \newblock \href {http://dx.doi.org/10.1016/j.astropartphys.2007.11.012}
  {\path{doi:10.1016/j.astropartphys.2007.11.012}}.

\bibitem{ArdouinBelletoileCharrier2005}
D.~{Ardouin}, A.~{Bell\'etoile}, D.~{Charrier}, {et al. (The CODALEMA Collaboration)},
%{Radio-detection signature of high-energy cosmic rays by the CODALEMA experiment},
  Nucl. Instr. Meth. A 555 (2005) 148--163.
  \newblock \href {http://dx.doi.org/10.1016/j.nima.2005.08.096}
  {\path{doi:10.1016/j.nima.2005.08.096}}.

\bibitem{FalckeNature2005}
H.~{Falcke}, W.~D. {Apel}, A.~F. {Badea}, {et al. (The LOPES Collaboration)},
%{Detection and imaging of atmospheric radio flashes from cosmic ray air showers},
  Nature 435 (2005)
  313--316.
  \newblock \href {http://dx.doi.org/10.1038/nature03614}
  {\path{doi:10.1038/nature03614}}.

\bibitem{LautridouArena2008}
P.~{Lautridou for the CODALEMA Collaboration},
%{},
  in: Proc. of the ARENA 2008 Conference, Rome, Italy, NIM A, 2009.
  \newblock \href {http://dx.doi.org/10.1016/j.nima.2009.03.164}
  {\path{doi:10.1016/j.nima.2009.03.164}}.

\bibitem{HaungsArena2008}
A.~{Haungs}, W.~D.~{Apel}, J.~C.~{Arteaga et al. (The LOPES Collaboration)},
%{},
  in: Proc. of the ARENA 2008 Conference, Rome, Italy, NIM A, 2009.
  \newblock \href {http://dx.doi.org/10.1016/j.nima.2009.03.033}
  {\path{doi:10.1016/j.nima.2009.03.033}}.

\bibitem{HuegeUlrichEngel2008}
T.~{Huege}, R.~{Ulrich}, R.~{Engel},
%{Dependence of geosynchrotron radio emission on the energy and depth of maximum of cosmic ray showers},
  Astroparticle Physics 30 (2008) 96--104.
  \newblock \href {http://dx.doi.org/10.1016/j.astropartphys.2008.07.003}
  {\path{doi:10.1016/j.astropartphys.2008.07.003}}.

\bibitem{CoppensArena2008}
J.~{Coppens for the Pierre Auger Collaboration},
%{Observation of radio signals from air showers at the Pierre Auger Observatory},
  in: Proc. of the ARENA 2008 Conference, Rome, Italy, NIM A, 2009.
  \newblock \href {http://dx.doi.org/10.1016/j.nima.2009.03.119}
  {\path{doi:10.1016/j.nima.2009.03.119}}.

\bibitem{FliescherArena2008}
S.~{Fliescher for the Pierre Auger Collaboration},
%{Radio detector array simulation: A full simulation chain for an array of antenna detectors},
  in: Proc. of the ARENA 2008 Conference, Rome, Italy, NIM A, 2009.
  \newblock \href {http://dx.doi.org/10.1016/j.nima.2009.03.068}
  {\path{doi:10.1016/j.nima.2009.03.068}}.

\bibitem{RevenuArena2008}
B.~{Revenu for the Pierre Auger Collaboration},
%{Radiodetection of cosmic air showers with autonomous radio detectors installed at the Pierre Auger Observatory },
  in: Proc. of the ARENA 2008 Conference, Rome, Italy, NIM A, 2009.
  \newblock \href {http://dx.doi.org/10.1016/j.nima.2009.03.028}
  {\path{doi:10.1016/j.nima.2009.03.028}}.

\bibitem{vandenBergIcrc2009}
A.~{van den Berg for the Pierre Auger Collaboration},
%{Radio detection of high-energy cosmic rays at the Pierre Auger Observatory},
  in: Proc. of the 31st ICRC, Lodz, Poland, 2009, {\em in press}.

\bibitem{KlagesICRC2007}
H.~O. {Klages for the Pierre Auger Collaboration},
%{HEAT - Enhancement Telescopes for the Pierre Auger Southern Observatory},
  in: Proc. of the 30th ICRC, Merida, Mexico, Vol.~5, 2008, pp. 849--852.

\bibitem{EtchegoyenIcrc2007}
A.~{Etchegoyen for the Pierre Auger Collaboration},
%{AMIGA, Auger Muons and Infill for the ground array},
  in: Proc. of the 30th ICRC, Merida, Mexico, Vol.~5, 2008, pp. 1191--1194.

\bibitem{HuegeUlrichEngel2007a}
T.~{Huege}, R.~{Ulrich}, R.~{Engel}, Astropart. Physics 27 (2007) 392--405.
  \newblock \href {http://dx.doi.org/10.1016/j.astropartphys.2007.01.006}
  {\path{doi:10.1016/j.astropartphys.2007.01.006}}.

\end{thebibliography}
%\bibliographystyle{elsarticle-num}



%%\end{thebibliography}

\end{document}